\documentclass[preprint,showpacs,preprintnumbers,amsmath,amssymb]{revtex4}
\usepackage{graphicx}
\usepackage{color}
\usepackage{pifont}

\def\bea{\begin{eqnarray}}
\def\eea{\end{eqnarray}}

\def\p{\partial} 
\def\nn{\nonumber}

\def\rt{\rightarrow}

\def\la{\langle}
\def\ra{\rangle}

\def\om{\omega}
\def\Om{\Omega}

\def\f{\frac}
\def\bJ{\bar{J}}

\begin{document}

\title{Particle current in a symmetric exclusion process with
  time-dependent hopping rates}
\author{Rahul Marathe$^1$, Kavita Jain$^2$ and  Abhishek Dhar$^1$}
\email{rahul@rri.res.in, jain@jncasr.ac.in, dabhi@rri.res.in}
\affiliation{$^1$Raman Research Institute, Bangalore 560080}
\affiliation{$^2$Theoretical Sciences Unit, Jawaharlal Nehru Centre
  for Advanced Scientific Research, Jakkur P.O., Bangalore 560064,
  India}
\date{\today} 
\begin{abstract}
In a recent study (Jain et al 2007 Phys. Rev. Lett. 99 190601), a
symmetric exclusion process with time-dependent 
hopping rates was introduced. Using simulations and a perturbation
theory, it was shown that if the hopping rates at two
neighboring sites of a closed ring vary periodically in time and have a
relative phase difference, there is a net $DC$ current which 
decreases inversely with the system size. In this work, we simplify
and generalize our earlier treatment. 
We study a model where hopping rates at all sites
vary periodically in time, and show that for certain choices of relative
phases, a $DC$  
current of order unity can be obtained. Our results are obtained using 
a perturbation theory in the amplitude of the time-dependent part of
the hopping  rate. We also present results obtained in a sudden
approximation that assumes large modulation frequency. 
\end{abstract}

\maketitle


\section{introduction}

The symmetric exclusion process (SEP) is one of the simplest and well
studied models of a stochastic interacting particle
system. 
In this model which can be defined on a $d$-dimensional hypercubic
lattice, particles 
move diffusively while satisfying the hard core constraint that two
particles cannot be on the 
same site. A  number of exact results have been obtained for this
model, particularly in one dimension \cite{Liggett:1985,Zia,Schutz:2000}. If the model is defined on a ring and
conserves the total density, the system obeys the equilibrium
condition of detailed balance in the steady state and thus does not
support any net current. A lot of attention has also been given to non-equilibrium 
steady states of driven SEP in which the particles can
enter or leave the bulk at the boundaries. For this model, the 
time-dependent correlation functions \cite{Santos:2001} and dynamical
exponents have been obtained  using the equivalence of the transition
matrix ($W$-matrix) to the Heisenberg model \cite{SchutzHeisenberg:1995}. 
Recently, large deviation functional and  
current fluctuations have also been calculated for the driven SEP 
\cite{Spohn:1983,Derrida:2002,Derrida:2004}.  Experimentally it has been shown that
SEP  can be used to model the diffusion of colloidal particles in
narrow pores
\cite{GuptaSEPexp:1995,HahnSEPexp:1996,KuklaScienceZeolite:1996,Wei:2000,Chou:1998,Chou:1999}.

Motivated by studies on quantum pumps where oscillating voltages
can drive electron current across a wire
\cite{Thouless:1983,Buttiker:1994,Aleiner:1998,Brouwer:98,Switkes:1999,Altshuler:1999,
Sharma:01,Ahorny:02,Watson:2003,Andrei:03,Leek:05,Strass:05,Sela:06,Diptiman:07},
we have recently shown that similar effect can occur in a SEP model in
which the hopping rates at  two neighbouring sites are chosen to vary
periodically in time  
and with a relative phase difference  \cite{jain2007}. 
Our  results obtained using Monte-Carlo simulations 
and a second-order perturbative calculation in the amplitude $f_1$ of the
time-dependent part of the hopping rate can be summarized as
follows: (i)  A $DC$ current $\bJ$ is obtained, which decays with system size $L$ as $\bJ \sim
1/L$. Correspondingly the time averaged density
profile varies linearly in the bulk of the system. 
(ii) The {\it DC} current $\bJ$ depends 
sinusoidally on the phase difference between rates at two sites. 
(iii) The dependence of $\bJ$ on driving frequency $\om$ shows a peak
at a frequency $\om^{*}$ with $\bJ \rt 1/\om$ as  
$\om\rt \infty$ and $\bJ \rt \om$ as $\om\rt 0$. The latter result means that a finite
number of particles are circulated even in the adiabatic limit.

Classical pumping of particles and heat, in similar time-dependent stochastic models,
has also been studied in
\cite{Astumian:2001,marathe07,sinitsyn07,ohkubo08} and seen in experiments
\cite{Astumian:2003}. Systems exhibiting pumping effect have often been modeled as Brownian
ratchets in which non-interacting particles move in an external
periodic potential. 
As discussed in \cite{astumian02}, these pump models are similar
to  Brownian ratchets \cite{Reimann:2002} where non-interacting
particles placed in
spatially asymmetric  potentials that vary periodically in time and
acted upon by noise execute directed motion. 
Models of non-interacting particles moving in symmetric potentials
have also been considered \cite{parrondo98,usmani02} and pumping demonstrated.
However for the model
studied by us, particle interactions seem necessary for  the pumping 
effect . Our model differs from such models in that  
here we are dealing with a many body particle system with
interactions. For such an extended system, as described in the following section, the $n$-point equation does not
close and involves next order correlation functions also.  
Pumping effect has also been found in the steady state of a driven SEP with
two species ($A$ and $B$) of particles in which both species have the same
diffusion constant \cite{Schutz2species:07}. In this case, although
the total current $J_A+J_B$ due to both species obeys the Fick's law,
the current due to one of the species does not follow the density
gradient. However the pumping mechanism is different from that in our
model where it arises due to the time-dependent rates.

In this paper, we  consider a generalization of our earlier
model by allowing the rates at all the sites to be time-dependent with
a relative phase difference between neighbouring sites.  The model
is treated analytically using two 
approximations: a perturbation theory in the time-dependent part of
the driving, and an 
expansion in the large frequency limit to leading orders. The
treatment in this paper considerably simplifies the earlier one given
in \cite{jain2007}.
The most interesting new result is that 
 in the model with time-dependent rates at all sites, a
current of order unity can be obtained even in the thermodynamic limit
for certain choices of relative phase differences.

The  paper is organized as follows. In section
(\ref{moddef}), the model is defined. In section~(\ref{purturb}) the
details of the first perturbation theory (expansion in $f_1$) are given, and two special choices
of hopping rates are discussed.  
The results obtained from a sudden approximation (expansion in
$1/\om)$) are given in section~(\ref{sudden}). Finally we end with a
discussion in section~(\ref{discuss}).


\section{Definition of Model}\label{moddef}

The model is defined on a ring with $L$ sites. A site $l=1,2,3,...L$
can be occupied by $n_l = 0$ or $1$ 
particle and the system contains a total of $N = \rho L$ particles where 
$\rho$ is the total density. 
A particle at site $l$ hops to an empty site either on the left or
right  with equal rates given by:
\bea
u_l &=& f_0 + f_1 v_l \nn \\ 
{\rm where}~~ v_l &=& \alpha_l \sin (\om t +\phi_l) = \nu_l e^{i \om t}+
\nu_l^* e^{-i \om t}~.
\eea
Here the site-dependent complex amplitudes are defined by $\nu_l =
\alpha_l~e^{i\phi_{l}}/2i$ with $\alpha_l$ real and 
 $f_1$ is chosen such that all
hopping rates are positive. We will discuss two particular choices for 
the hopping rates in detail. Our first choice corresponds to the
case where the hopping rates are time-dependent at only two sites of the
ring, and we get an average current which decays inversely with
system size. In the second case, we choose time-dependent hopping rates at
all sites and show that a finite current can be
obtained even in the thermodynamic limit.

A configuration of the system can be  specified by the set $\{ n_l
 \}$, $l=1,2,...L$. Let us define ${\bf P}(t)$ as the probability vector
in the configuration space, with elements $P(C,t)$ giving the
 probability of the system being  in the configuration $C=\{n_l\}$ at time
 $t$. Then the stochastic dynamics of  the many particle system is 
described by the master equation:
\bea
\frac{d {\bf P}(t)}{dt}= {\bf W}(t) ~{\bf P}(t)= {\bf W_0} ~{\bf
  P}(t)+ {\bf W_1}(t) ~{\bf P}(t)
\label{master}
\eea
where ${\bf W}$ is the transition matrix, which we
have split into a time-independent and a time-dependent part. 
One can also consider the 
time-evolution equations for $m$-point equal-time correlation functions
$C_{l_1,l_2,l_3,....,l_m}(t)= \langle n_{l_1} ... n_{l_m}
\rangle=\sum_{\{n_l\}} n_{l_1}...n_{l_m} P(\{n_l\},t)$. 
Thus,
for example, the density $\rho_l(t)=\la n_l\ra$ and the two-point
correlation function $C_{l,m}(t)$ satisfy the following equations:
\bea
&&\f{\p \rho_l}{\p t}\ + 2 u_l \rho_l - u_{l-1} \rho_{l-1} - u_{l+1} \rho_{l+1}
= u_l ( C_{l-1,l} + C_{l,l+1}) - u_{l+1} C_{l,l+1} - u_{l-1} C_{l-1,l}
\label{denseq} \\
&& \frac{\partial C_{l,m}}{\partial t}\ + 2(~u_l +
u_m~)C_{l,m}-u_{l-1}~C_{l-1,m}-u_{l+1}~C_{l+1,m} 
-u_{m-1}~C_{l,m-1}-u_{m+1}~C_{l,m+1}\nn\\
&=& u_l~(~C_{l-1,l,m} + C_{l,l+1,m}~)+u_m~(~C_{l,m-1,m} + C_{l,m,m+1}~)
-
u_{l-1}~C_{l-1,l,m}-u_{l+1}~C_{l,l+1,m}\nn\\
&&-u_{m-1}~C_{l,m-1,m}-u_{m+1}~C_{l,m,m+1}, ~~{\rm for}~~~ |l-m|\neq1  \nn \\
&& \frac{\partial C_{l,l+1}}{\partial t}\ + (~u_l +
u_{l+1}~)C_{l,l+1}-u_{l-1}~C_{l-1,l+1}
-u_{l+2}~C_{l,l+2}\nn\\
&=& u_l~~C_{l-1,l,m} +u_{l+1}~ C_{l,l+1,l+2}~
-u_{l-1}~C_{l-1,l,l+1}-u_{l+2}~C_{l,l+1,l+2}~. 
\label{correleq}
\eea

From Floquet's theorem \cite{jung93}, it follows that the long time
state of the system (assumed to be unique)  
will be periodic in time, with period $T=2 \pi/\omega$. Here we will 
be mainly interested in the $DC$ current ${\bar J}$ defined as 
\bea
\bJ_l = \f{1}{T}\ \int_{0}^{T} J_{l,l+1}(t)~dt,
\eea
where the current $J_{l,l+1}$ in a bond connecting sites $l$ and $l+1$ is given by 
\bea
J_{l,l+1} = u_l(\rho_l - C_{l,l+1}) - u_{l+1} ( \rho_{l+1} -
C_{l,l+1})
\label{bondJ}
\eea
and the local density $\rho_l= \langle n_l \rangle$. 
From the periodicity of the state at long times and particle conservation, it
follows that the $DC$ current is uniform in space and
therefore, using Eq.~(\ref{bondJ}), we can write for the $DC$ current:
\bea
\bJ &=& \f{1}{L T} \int_{0}^{T} \sum_{l=1}^L J_{l,l+1}(t)~dt \\
   &=& \f{f_1}{L T} \int_{0}^{T} \sum_{l=1}^L (v_{l+1}(t)-v_l(t))C_{l,l+1}(t)
~dt 
\label{Jdc}
\eea
Thus to find the $DC$ current,  we need to compute the two-point
correlation function $C_{l,l+1}(t)$. 
In this paper, we will develop two different  perturbation schemes,
valid for general $v_l$,
and then apply them to some special choices of the rates $v_l$.

Note that for $f_1 = 0$, the above model reduces to the homogeneous SEP with
periodic boundary conditions whose properties
are known exactly. In this case the steady state is an equilibrium state which obeys detailed balance and
hence the average current is zero (note that this result holds even when the rates
$\{u_l\}$ are site-dependent, but time-independent).
In the steady state,  all configurations are
equally probable i.e. $P(C)=1/{L \choose N}$ when $f_1=0$. Then one
can show that the density and correlation functions for the homogeneous 
SEP are given by:
\bea
\rho_l^{(0)}&=&\rho = \f{N}{L}\ \nn\\
C_{l_1,l_2}^{(0)}&=& \rho \f{(N-1)}{(L-1)}\ \nn\\
C_{l_1,l_2,l_3,....,l_m}^{(0)} &=& {L-m \choose N-m}/{L \choose N}.
\label{SEPcorr}
\eea
\section{Perturbation theory: expansion in $f_1$.}
\label{purturb}

For $f_1 \neq 0$, the knowledge of the exact steady state of
homogeneous SEP enables us to set up a 
perturbation expansion
in $f_1$ of various observables. We now describe this perturbation
theory within which we calculate an expression for ${DC}$ current 
$\bJ$ in the bulk of the system. A similar perturbation technique was
developed for a two-state system in \cite{astumian89}. 
We  expand various quantities of interest  
with $f_1$ as the perturbation parameter about the homogeneous steady
state corresponding to $f_1 = 0$.  
Thus we write
\bea 
\rho_l(t) &=& \la n_l(t) \ra = \rho + \sum_{r=1}^{\infty} f_1^{r}
\rho_l^{(r)}(t)\label{Rhopert}  \\
C_{l,m}(t) &=& \la n_l(t)n_m(t) \ra = C_{l,m}^{(0)} +
\sum_{r=1}^{\infty} f_1^{r} C_{l,m}^{(r)}(t)~,\label{Cpert} 
\eea
and similar expressions for higher correlations. 
Plugging in  Eq.~(\ref{Cpert}) into Eq.~(\ref{Jdc}), we find that the
lowest order contribution to $\bar{J}$ is at ${\cal{O}}(f_1^2)$ and
given by:  
\bea
\bJ^{(2)} = \f{f_1^2}{T~L} ~\int_{0}^{T} \sum_{l=1}^L (~v_l~-~v_{l+1}~)~
C_{l,l+1}^{(1)}~dt~.
\label{Jlast}
\eea
To develop our perturbation theory and find two-point correlation function $C_{l,m}^{(1)}$,
we start with  the time evolution equation for density $\rho_l(t)$
which is given by Eq.~(\ref{denseq}). 
Plugging in the expansions in Eqs.(\ref{Rhopert}) and (\ref{Cpert}),
we get the following equation for 
the density $\rho_{l}^{(r)}$ at $r^{th}$ order: 
\bea
\f{\p \rho_{l}^{(r)}}{\p t}\ &-& f_0 \Delta_{l}\rho_{l}^{(r)} + 2 v_l \rho_{l}^{(r-1)} - v_{l-1} 
\rho_{l-1}^{(r-1)} - v_{l+1} \rho_{l+1}^{(r-1)}\nn\\
 &=& v_l ( C_{l-1,l}^{(r-1)} + C_{l,l+1}^{(r-1)}) - v_{l-1} C_{l-1,l}^{(r-1)} - v_{l+1} C_{l,l+1}^{(r-1)},
\label{rhok}
\eea
where $\Delta_l g_l = g_{l+1} - 2 g_l + g_{l-1}$ defines the discrete Laplacian operator. Thus the density 
at $r^{th}$ order can be obtained in terms of density and two point correlation function at $(r-1)^{th}$
order. We check  that at the zeroth order, we obtain the homogeneous SEP for
which  the density and 
all equal time correlations are given by Eq.~(\ref{SEPcorr}). 
At first order, the above equation then gives:
\bea
\f{\p \rho_l^{(1)}}{\p t}\ - f_0 \Delta_{l}\rho_{l}^{(1)} = r_0 \Delta_{l} v_l,\label{rhol}
\eea
where $r_0 = \rho - C_{l,m}^{(0)}$. The solution for this equation is the sum of a homogeneous part
which depends on initial conditions and a particular integral. At long times the homogeneous part 
vanishes while the particular integral has the following asymptotic form:
\bea
\rho_{l}^{(1)}(t) = A_{l}^{(1)}e^{i\om t} + A_{l}^{*(1)}e^{-i\om t}.\label{rho1_t}
\eea
Substituting Eq.~(\ref{rho1_t}) in Eq.~(\ref{rhol}) we obtain the following equation for 
$\{ A_{l}^{(1)}\}$:
\bea
(i\om + 2 f_0)A_{l}^{(1)}- f_0 A_{l-1}^{(1)}- f_0 A_{l+1}^{(1)} = r_0 ( \nu_{l+1} - 2\nu_{l} + \nu_{l-1} )~.  
\label{Arho1}
\eea
This can be written in matrix form as:
\bea
{\hat Z}(\om)~~ {\bf A} = -r_0 ~{\hat B}~{\bf \Phi },
\eea
where
\bea
Z_{lm} &=& -f_0~ \delta_{l,m+1} + (~i\om +2f_0~)~\delta_{l,m}-f_0~\delta_{l,m-1} \nn\\
B_{lm} &=& - \delta_{l,m+1} + 2~\delta_{l,m}- \delta_{l,m-1}\nn\\
{\bf A} &=& \{A_1^{(1)},A_2^{(1)},....,A_L^{(1)}\}^T, {\bf \Phi} = \{\nu_1,\nu_2,....,\nu_L\}^{T},
\label{matB}
\eea
and periodic boundary conditions are implicitly taken. The above equation can be solved for ${\bf A}$ 
and we get:
\bea
{\bf A} = -{r_0} ~{\hat G}(\om)~{\hat B}~{\bf \Phi },
\eea
where ${\hat G}(\om) = {\hat Z^{-1}}(\om )$. Both ${\hat G}(\om)$ and
${\hat B}$ are cyclic matrices and 
so can be diagonalized simultaneously. The eigenvalues of ${\hat Z}(\om )$
are $i\om + 4 f_0 \sin^{2}(p\pi /L)$, while that of ${\hat B}$ are $4
\sin^{2}(p \pi /L)$ with $p = 1,2,...,L$, and eigenvector elements are
$e^{i 2\pi p l/L}/L^{1/2}$.  
Hence $A_{l}^{(1)}$ can be written as:
\bea
A_{l}^{(1)} = -\f{4r_0}{L}\ \sum_{m=1}^{L}\sum_{p=1}^{L}~~ 
\f{e^{-i\f{2\pi p (l-m)}{L}}~\sin^{2}(p\pi/L)~}{i\om + 4f_0
\sin^{2}(p\pi/L)}\ \nu_m, \label{A_k1s} 
\eea
which in the large $L$ limit gives:
\bea
A_{l}^{(1)}= -\f{r_0}{f_0}\ \nu_l ~+~ \f{ir_0\om}{f_0^2}\ \f{1}{z_+-z_-}
\sum_{m=1}^{L}~ [~z_-^{|m-l|}~+~z_-^{L-{|m-l|}}~]~\nu_m ,\label{AlLlarge}
\eea
where, $z_-=y/2 - [(y/2)^2-1]^{1/2},~z_+=1/z_-$ and $y=2+({i\om}/{f_0})$.

To compute the 
${\cal{O}}(f_1^2)$ contribution to $\bar{J}$, we need to evaluate $C_{l,m}^{(1)}$,
which we now proceed to obtain.  
Inserting the perturbation series in Eqs.~(\ref{Rhopert}) and
(\ref{Cpert}) into Eq.~(\ref{correleq}) we get the following equation
for the correlation $C_{l,m}^{(r)}$ at $r^{th}$ order for $|m-l|\ne 1$:
\bea
\frac{\partial C_{l,m}^{(r)}}{\partial t}\ &-& f_0~(~\Delta_l+\Delta_m~)~C_{l,m}^{(r)}
+2v_l~C_{l,m}^{(r-1)}-v_{l-1}~C_{l-1,m}^{(r-1)}-v_{l+1}~C_{l+1,m}^{(r-1)}\nn\\
&+&2v_m~C_{l,m}^{(r-1)}-v_{m-1}~C_{l,m-1}^{(r-1)}-v_{m+1}~C_{l,m+1}^{(r-1)}\nn\\
&=& v_l~(~C_{l-1,l,m}^{(r-1)} + C_{l,l+1,m}^{(r-1)}~)+v_m~(~C_{l,m-1,m}^{(r-1)}+ C_{l,m,m+1}^{(r-1)}~)\nn\\
&-&v_{l-1}~C_{l-1,l,m}^{(r-1)}-v_{l+1}~C_{l,l+1,m}^{(r-1)}-v_{m-1}~C_{l,m-1,m}^{(r-1)}-
v_{m+1}~C_{l,m,m+1}^{(r-1)},\nn\\
{\rm while ~ for}~~~ m = l+1:\nn\\
\frac{\partial C_{l,l+1}^{(r)}}{\partial t}\ &+&f_0~(~2C_{l,l+1}^{(r)}-C_{l-1,l+1}^{(r)}-C_{l,l+2}^{(r)}~)\nn\\
&=&v_{l+2}~(~C_{l,l+2}^{(r-1)}~-~C_{l,l+1,l+2}^{(r-1)}~)~+~v_{l-1}~(~C_{l-1,l+1}^{(r-1)}~-~C_{l-1,l,l+1}^{(r-1)}~)\nn\\
~&-&~v_{l}~(~C_{l,l+1}^{(r-1)}~-~C_{l-1,l,l+1}^{(r-1)}~)~-~v_{l+1}~(~C_{l,l+1}^{(r-1)}~-~C_{l,l+1,l+2}^{(r-1)}~).
\label{Clmevol}
\eea

At first order we get: 
\bea
\frac{\partial C_{l,m}^{(1)}}{\partial t}\ &-& f_0(\Delta_l +\Delta_m) C_{l,m}^{(1)} =
 k_0(\Delta_l v_l + \Delta_m v_m)~,\nn\\
\frac{\partial C_{l,l+ 1}^{(1)}}{\partial t}&+& f_0 \left( 2 C_{l,l + 1}^{(1)}
 -C_{l-1,l+1}^{(1)}-C_{l,l + 2}^{(1)} \right) 
= k_0(v_{l-1}+v_{l+2}-v_l-v_{l+1}),  
\label{corr-all}
\eea 
where $k_{0}= C_{l_1,l_2}^{(0)}- C_{l_1,l_2,l_3}^{(0)} $ and these are
known from Eq.~(\ref{SEPcorr}). 
The computation of even the homogeneous solution 
of the above set of equations is in general a non-trivial task because of the form of the equations
involving nearest neighbor indices and requires a Bethe ansatz or dynamic product ansatz 
\cite{Schutz:2000,Santos:2001}. However it turns out that the long time solution can still 
be found exactly and is given by:
\bea
C_{l,m}^{(1)}(t)=\frac{k_0}{r_0} [\rho_l^{(1)}(t)+\rho_m^{(1)}(t)]= A_{l,m}^{(1)} e^{i \omega t} + 
A_{l,m}^{*(1)} e^{-i \omega t}~,\label{corr-soln}
\eea
where $A_{l,m}^{(1)}=(k_0/r_0)(A_l^{(1)}+A_m^{(1)})$. 
It is easily verified that this satisfies Eq.~(\ref{corr-all}) for all 
$l, m$. To determine whether the system indeed has a product measure 
requires a  more detailed analysis of the higher order terms in the perturbation series 
and higher correlations. We have verified that at least to first order in
perturbation theory,  all correlation functions in fact have the same
structure as the two-point correlation function in Eq.~(\ref{corr-soln}).

We now plug the solution in Eq.~(\ref{corr-soln}) into
Eq.~(\ref{Jlast}) for the average current in the system and after some simplifications obtain:
\bea
\bar{J}^{(2)} = -\f{f_1^2}{L}\ ~\frac{k_0}{r_0}~\sum_{l=1}^{L}~(~A_{l+1}^{*(1)} \nu_{l}~+~A_{l+1}^{(1)} \nu_{l}^{*}~-
~A_{l}^{*(1)} \nu_{l+1}~-~A_{l}^{(1)} \nu_{l+1}^{*}~)~,\label{Ja_lnu_ls}
\eea
with $A_l^{(1)}$ given by Eq.~(\ref{AlLlarge}). For any given choice of the
rates $\nu_l$, this general expression can be used to explicitly evaluate the
net $DC$ current in the system.  
\begin{figure}[t]
\includegraphics[width=5in]{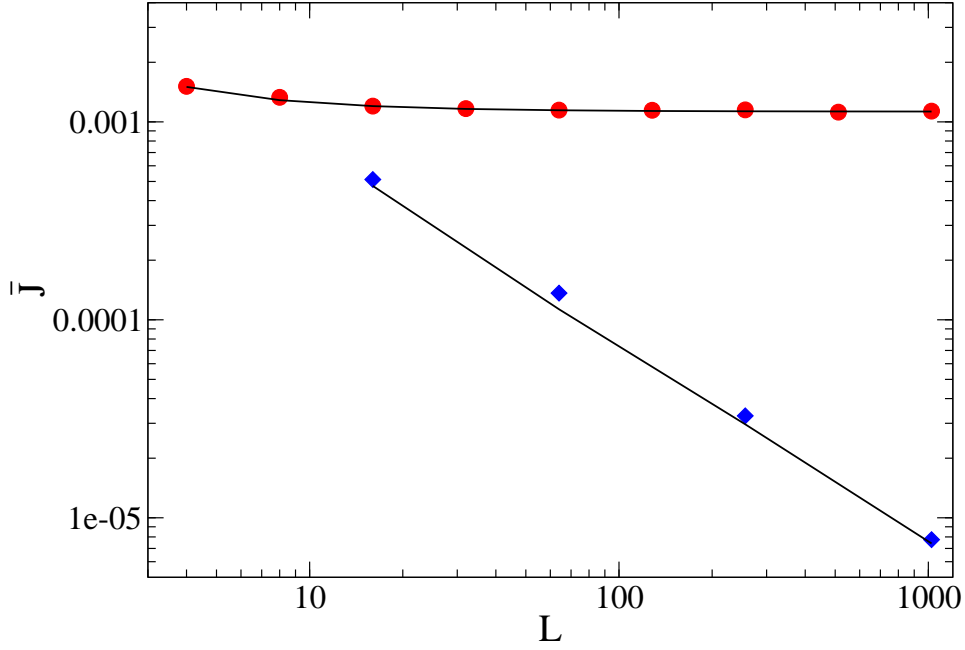}
\caption{Log-log plot of $\bJ$ versus system size $L$ at half filling
 for two cases discussed in the text. 
The points ({\color{blue}\ding{117}}) correspond to simulations for
the $2$ special sites problem with $\phi=\pi /2$,  
$f_0 =0.3$, $f_1=0.2,~\omega=0.2 \pi$. In this case the current
goes as $\sim L^{-1}$. The points ({\color{red}\ding{108}}) corresponds to all sites having
time-dependent hopping rates with  $q=\pi/2$ and 
$f_0=0.5,~f_1 =0.1,~\omega=0.2 \pi$. In this case 
the current  goes to a constant value at large
$L$. The bold lines indicate the results from the perturbation theory.}
\label{JvsLall}
\end{figure}
\begin{figure}
\includegraphics[width=6in]{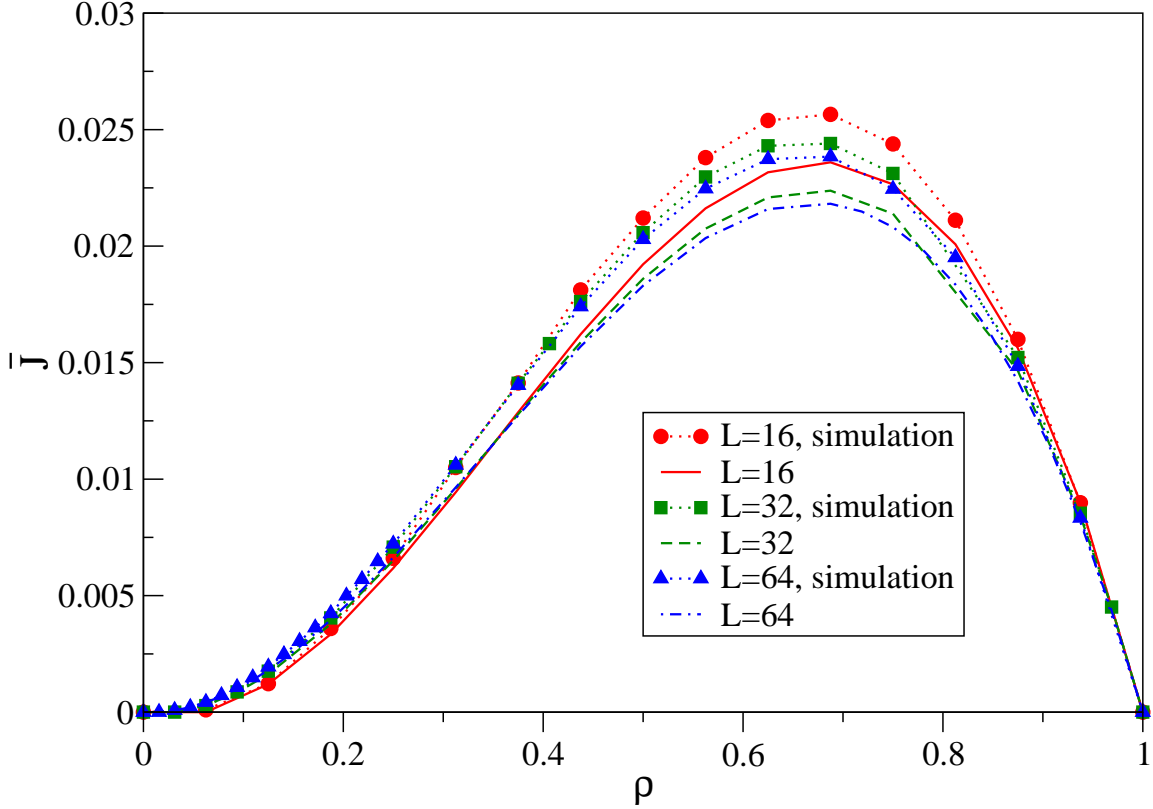}
\caption{Plot of ${DC}$ current $\bJ$ versus density $\rho = N/L$ for parameters $f_0=0.5$,
$f_1=0.4$, $\om = 0.2 \pi$ and $\phi_l =\pi l/2$ for system sizes
$L=16, 32$ and $64$. 
Both the results from simulations (symbols connected by dotted lines)
and from the perturbation theory (lines) are plotted.}
\label{Jvsrho}
\end{figure}
We now consider two special choices of the rates $\{\nu_l\}$.

\noindent{\bf (i)} The choice $\alpha_1=\alpha_L=1$, all other
$\alpha_l=0$, and $\phi_1=0,\phi_L=\phi$ corresponds to the 
pumping problem with two special sites studied in \cite{jain2007}.    
In the limit of large $L$, this gives:
\bea
{\bar J}^{(2)}= \left(\frac{f_1}{f_0}\right)^2  \frac{k_0 \omega \sin
  \phi}{L}~ {\rm Re}[z_-], \label{j2site}
\eea
which agrees with the result presented in \cite{jain2007} ( apart from
a factor of two which was missed in that paper). Writing $z_+=r e^{i
\eta}$, we find that for $\omega \ll \omega^*=2 f_0$, the magnitude $r
\approx 1+\sqrt{\omega/\omega^*}$ and the angle $\eta \approx
\sqrt{\omega/\omega^*}$. In the opposite limit, $r \approx 2
\omega/\omega^*$ and $\eta \approx \pi/2-\omega^*/\omega$. Using
$z_+=1/z_-$, we find that   
the current has the scaling form:
\bea
{\bar J}^{(2)} =\frac{f_1^2 k_0\sin \phi}{f_0 L}
G\left(\frac{\omega}{2 f_0} \right)
\label{scaling}
\eea
where the scaling function $G(x)=2 x$ for $x \ll 1$ and $1/x$ for $x
\gg 1$. We note that $\bar{J}$ is independent of $f_0$ for large
$x$. This can be seen by writing the master equation as:
\bea
\frac{d {\bf P}}{d (\omega t)}= \frac{f_0}{\omega} {\bf W_0 P(t)}+
\frac{f_1}{\omega} {\bf W_1 P(t)}~. 
\eea
For $\omega \gg f_0$, the first term on
the right hand side can be neglected thus giving the probability
distribution to be a function of $f_1/\omega$.

\noindent{\bf (ii)} The second case we consider here assumes 
$\alpha_l=1$ at all sites and $\phi_l=ql$, where $q=2\pi s/L$ with
 $s=1,2...L/2$, so that there is a constant phase difference $q$
 between successive sites.  
 In this case, $A_{l}^{(1)}$'s given by Eq.~(\ref{A_k1s}), evaluated at
 large $L$ gives: 
\bea
A_l^{(1)}&=&\f{i r_0}{2 f_0}e^{i q l} a \\
{\rm where}~~a&=&\f{1-\cos q}{y/2-\cos q} \nn 
\eea
and from Eq.~(\ref{Ja_lnu_ls}) we get for the average current:
\bea
\bar{J}^{(2)}&=&-\f{f_1^2 k_0}{f_0}\sin q ~{\rm Im} [a]~ \nn\\
&=& \f{2~f_1^2~ k_0~ \om~ \sin q~ (1-\cos q)}{[~\om^2~+~4f_0^2~(1-\cos q)^2~]}\ . \label{jallsites} 
\eea
Thus we see that for most values of $q$ we get a finite current, even
in the limit $L \to \infty$. For $q \sim 1/L$ and $q \sim \pi -1/L$,
the current goes to zero for large system size as $\bar{J} \sim L^{-3}$. 
From the current expression in Eq.~(\ref{jallsites}), we can find out
the value  $q=q^*$, at which
the current is a maximum. By differentiating Eq.~(\ref{jallsites}) with respect to $q$ we get: 
\bea
\cos(q^*)=(1+\Om^2)-\sqrt{(1+\Om^2)^2- (1-\Om^2)},
\eea
where $\Om =\om/2f_0$. It turns out that for large $\om$ the maximum is at $q^*=2\pi/3$,  
while for small frequencies  we get $q^* \sim \sqrt{\om}$. 
Also we find from Eq.~(\ref{jallsites}) that in the adiabatic and fast drive limits,
the currents are respectively given by:
\bea
 \bar{J}^{(2)}= \begin{cases}  \f{f_1^2 k_0}{2f_0^2}~{\cot (q/2)}~
 \om ~~~,~~~\om/f_0 << (1-\cos q) \\
 2f_1^2 k_0 \sin q (1-\cos q)~\f{1}{\om}~~~,~~~ \om/f_0 >>1~. \label{jlimits} 
\end{cases}
\eea

The perturbation theory results described above turn out to be quite accurate as can
be seen from the comparisons with simulation results shown in
Fig.~(\ref{JvsLall}) for both cases (i) and (ii).  In this figure, we
have plotted the current for different system sizes and verify the
$\bar{J} \sim L^{-1}$ dependence for case (i) and $\bar{J} \sim L^0$
for case (ii) with $q=\pi/2$. 
Using the expression for $k_0$ in Eqs.~(\ref{j2site}, \ref{jallsites}),
we find that ${\bar J}^{(2)} \sim \rho^2 (1-\rho)$  which has a  maximum at $\rho^*=2/3$ and breaks
particle-hole symmetry.  This particle-hole asymmetry can be
understood easily. 
From the definition of the model we see that, unlike the particles, the hopping
rates of a hole are not symmetric: a hole at site $l$ hops towards
right with rate $u_{l+1}$ and left with $u_{l-1}$. In
Fig.~(\ref{Jvsrho}) we have plotted simulation results for the average
current as a function of particle density, for different system sizes, and find
good agreement with our perturbative result, even at a relatively
large value of $f_1/f_0$.  

In simulations we have looked at the density profiles and find  that the site wise density
profile $\bar{\rho}_{l}$ in case (ii) is flat. This is unlike in
\cite{jain2007}, where we found high densities 
at the two special sites and then a linear density profile in the bulk. The flat density
profile, for case (ii), is understood because here there are no special \emph{pumping} 
sites. It is interesting that we can get current in the system even in the absence of Fick's law. 
We also note that even if the hop-out rates are made biased in one
direction, like in the asymmetric  exclusion process (ASEP), we can
still get a current opposing this bias (for small biases).

\section{Sudden approximation: $\omega/f_0 \gg 1$ }
\label{sudden}

In this section, we find the $DC$ current within sudden approximation 
 following the procedure of \cite{Reimann:1996}. Calling $\theta=\omega
t$, the master equation Eq.~(\ref{master}) can be rewritten as 
\bea
\frac{d {\bf P}(\theta)}{d \theta}= \frac{1}{\omega}
\left[{\bf W_0}+{\bf W_1}(\theta) \right] {\bf P}(\theta)
\eea
which can be expanded in powers of $1/\omega$ by using 
${\bf P}(\theta)= \sum_{n=0}^{\infty} \omega^{-n}
{\bf P}_s^{(n)}(\theta)$ to give
\bea
\frac{d {\bf P}_s^{(0)}}{d \theta} &=& 0 \\
\frac{d {\bf P}_s^{(1)}(\theta)}{d \theta} -{\bf W_1}(\theta) {\bf
  P}^{(0)}_s&=& {\bf W_0} {\bf P}^{(0)}_s
\label{sudden1}
\eea
and so on. From the zeroth order equation, we see that ${\bf P}^{(0)}_s$
is independent of $\theta$. In fact, 
for $\omega \to \infty$, we expect the system to behave as
the unperturbed homogeneous SEP for which ${\bf W_0} {\bf P}^{(0)}_s=0$
is satisfied and as discussed in Section \ref{moddef}, all the elements
of the vector ${\bf P}^{(0)}_s$ are known.  
Using this fact, the first order correction ${\bf P}^{(1)}_s$ can be found by 
integrating Eq.~(\ref{sudden1}) over $\theta$. Following steps as
those leading to Eq.~(\ref{Jlast}), we can obtain an expression for
average current 
$\bar{J}_s$ at order ${\cal{O}}(1/\om)$ which is given by:
\bea
{\bar J}_s^{(1)}= \frac{f_1}{2 \pi \omega L} \int_0^{2 \pi} d \theta~  \sum_{l=1}^L
(v_{l+1}-v_l)  {\tilde C}_{l,l+1}^{(1)} 
\eea
where we have expanded the nearest neighbor correlation function
$C_{l,l+1}= \sum_{n=0}^{\infty} \omega^{-n} {\tilde C}_{l,l+1}^{(n)}$ in powers
of $1/\omega$ and used the expression for ${\tilde C}_{l,l+1}^{(0)}=C_{l,l+1}^{(0)}$ given by
Eq.~(\ref{SEPcorr}). 
The first order correction to correlation function can be obtained by
perturbatively expanding Eq.~(\ref{correleq}) and obeys the following simple 
equation:  
\bea
\frac{d {\tilde C}^{(1)}_{l,l+1}}{d \theta} =f_1 k_0 ~(v_{l+2}+v_{l-1}-v_l-v_{l+1})~. 
\eea
We now again discuss the two special choices of rates $v_l$,
discussed in the previous section.

\noindent{\bf (i)} In this case, only two sites have time-dependent hopping rates.
Solving the equations above  for  the correlation function, we get:
\bea
{\tilde C}^{(1)}_{1,2} &=& f_1 k_0 (\cos (\theta)-\cos(\theta+\phi))+c_{1,2}\\
{\tilde C}^{(1)}_{L-1,L} &=& -f_1 k_0 (\cos (\theta)-\cos(\theta+\phi))+c_{L-1,L}\\
{\tilde C}^{(1)}_{L,1} &=& f_1 k_0 (\cos (\theta)+\cos(\theta+\phi))+c_{L,1}
\eea
where $c_{i,j}$ are constants of integration (which do not contribute to
current). Using the above equations in the expression for ${\bar J}^{(1)}_s$, 
we finally obtain
\bea
{\bar J}^{(1)}_s= \frac{2 f_1^2 k_0 \sin \phi}{\omega L}~.
\eea
Thus, we find that to leading order in $1/\omega$ (and arbitrary
$f_1$), 
the $DC$ current is the same as the one obtained by taking large $\omega$
limit in the current expression Eq.~(\ref{scaling}) obtained from the  $f_1$
expansion. 

\noindent{\bf (ii)} In this case with $\alpha_l=1$ at all sites, 
the equations for the first order correlation functions can be solved
for arbitrary phases $\phi_l$, and we get:
\bea
{\tilde C}_{l,l+1}^{(1)}=k_0 f_1
\left[\cos(\theta+\phi_l)+\cos(\theta+\phi_{l+1})-\cos(\theta+\phi_{l-1})\cos(\theta+\phi_{l+2})
  \right]~.
\eea
Using these in the current expression and after some simplifications, we get:
\bea
{\bar J}_s^{(1)}= \frac{k_0 f_1^2}{\omega L} \sum_{l=1}^L \left[2 \sin (\phi_{l+1}-\phi_l)-
\sin(\phi_{l+1}-\phi_{l-1}) \right].\label{Jdelphi}
\eea
Note that the above expression depends on the phase difference between
nearest and next nearest neighbor sites. For $\phi_l= q l$, we recover
the result stated in the second line of Eq.~(\ref{jlimits}).


\section{Discussion}\label{discuss}

In this article,  we have considered a lattice model of diffusing particles with hard core
interactions and shown that if the hopping rates at various sites are
chosen to be symmetric but time-dependent, a $DC$ current can be
generated in the system. Thus a ratchet effect is obtained in the
sense that a directed current occurs even though there is no net applied
external biasing force. Unlike many other examples of models of
classical ratchets, there is no asymmetric potential or asymmetric noise in our
model. However asymmetry is incorporated in the modulation of the
hopping rates, and this is best   seen when we consider the case where
the modulation is given by $v_l(t)=\sin ( \om t-ql)$. This of course
corresponds to a  wave  traveling in a \emph{given} direction. 
A non-trivial aspect of the problem studied is the fact that the
effect goes away as soon as we switch off the hard-core interactions.
For non-interacting particles, the $DC$ current given by 
${\bar J}= ({1}/{L T}) \int_0^T dt~ \sum_{l=1}^L u_l \rho_l- u_{l+1} \rho_{l+1}$,
is immediately seen to be exactly zero for arbitrary choice of the time-dependent
rates. On the other hand,  having 
interactions in the system is not a sufficient condition to generate a
$DC$ current. For the models considered in this paper,
the hopping rate is site-wise symmetric.
But if the hopping rates are symmetric bond-wise, {\it i.e.}, the hop
rate $u_{l,l+1}$ from site $l$ to $l+1$ is the same as that from $l+1$ to $l$,
then the  $DC$ current is zero for any choice of phases $\phi_l$. To see
this, consider the density evolution equation obeyed by bond-wise symmetric SEP: 
\bea
\frac{\partial \rho_l}{\partial t}=u_{l-1,l} (\rho_{l-1}-\rho_l)+u_{l,l+1} (\rho_{l+1}-\rho_l)
\eea
Unlike Eq.~(\ref{denseq}) for site-wise symmetric SEP, $\rho_l=\rho$ is
a solution of the above equation for any choice of rates $u_l$. In
fact, an inspection of the master equation shows that, even with a
time-dependent ${\bf W}$-matrix, all configurations are equally likely, thus
leading to zero current. Thus the exclusion process with bond-wise
symmetric rates does not give the ratchet effect. It is not completely
clear as to what are the necessary and sufficient conditions to get a
directed current \cite{sinitsyn08}.

For the model considered here, since the equations for any 
$n$-point 
correlation function do not close, it does not seem simple to solve
the model exactly. We have therefore studied the system analytically
using a perturbation theory in the amplitude $f_1$ and the inverse
frequency $1/\omega$. 
In this paper, we have been able to obtain the $DC$ current at order
$f_1^2$ by solving the evolution equations for density and two point
correlation function to order $f_1$. This is unlike our earlier
solution in \cite{jain2007} where the density was obtained to  second
order in $f_1$. Also, we have been able to obtain results for large
driving frequency by solving the correlation function alone by such
perturbative approaches. Comparing with simulations we find that the
perturbative results  turn out to be quite accurate. 

We now briefly comment on the adiabatic limit, which has been
much studied in the quantum context. In our case, from our
perturbation theory result we see that, over one time period of the
driving there is a finite particle 
transport, even in the adiabatic limit.   
Formally we can obtain an exact expression for the net particle transport. For 
this we start with the master equation 
${\p {\bf P}}/{\p t}= {\bf W}(t) {\bf P}$.
Let ${\bf P}^{(0)}(t)$ be the 
instantaneous equilibrium solution satisfying
${\bf W}(t) {\bf P}^{(0)}=0$.  
Then, for slow rates $\om$, ${\bf P}(t)$ will have the form 
${\bf P}^{(0)}(t)+\om {\bf P}^{(1)}(t)$ where the correction is given by:
$\om {\bf P}^{(1)}={\bf W}^{-1}~{\p {\bf P}^{(0)}}/{\p t}$~.
The net particle transported across any bond in one time cycle,
${\cal{N}}$, can then  be expressed as:  
\bea
{\cal{N}}
=\int_0^T dt \sum_C  J(C) P(C,t)
= -\int_0^{2 \pi} dx \sum_{C,C'} J(C)  \f{\p W^{-1}_{C,C'}(x)}{\p x} P^{(0)}(C',x)~,
\eea
where $J$ refers to the current on any given bond. 
Thus we have a formal expression, for  the net particle transported, in
terms of  an integral over an \emph {equilibrium average} of some 
quantity. However this expression does not appear to have any simple physical
interpretation and it is not easy to obtain  
any explicit results, unlike the fast case treated in section~(\ref{sudden}).
Recently adiabatic pumping phenomena have been studied in the context
of geometric phase interpretation \cite{sinitsyn07},
but the main focus has been on two-state stochastic systems.
In this case, the current from system to the reservoirs was
calculated using full counting statistic in the adiabatic or slow
driving regime.

Finally, we point out that an experimental realization of the effect
observed in our model should be possible in  colloidal systems.  
For instance, consider  a colloidal suspension in an externally applied
laser field. This constitutes a system of diffusive interacting
particles in an external potential (generated by the laser field) of
the form $V(x,t)=V_0 \sin (\om t - qx)$. This system is similar to the
model that we have studied. There are some differences, 
namely, in this case because the external field is space dependent,
hence the effective hopping rates are not symmetric in the forward and backward directions.
It would be  interesting to study this model to see if a current can be
generated here, and perhaps one can make detailed predictions for
experimental observation.

Acknowledgement: K.J. thanks T. Antal, K. Mallick,
A. Schadschneider and G.M. Sch{\"u}tz for useful discussions. R.M. thanks
A. Kundu for useful discussions.


\end{document}